\documentclass[12pt]{article}

\pdfoutput=1
\DeclareFontFamily{T1}{calligra}{}
\DeclareFontShape{T1}{calligra}{m}{n}{<->s*[1.44]callig15}{}
\DeclareMathAlphabet\mathcalligra   {T1}{calligra} {m} {n}
\DeclareMathAlphabet\mathzapf       {T1}{pzc} {mb} {it}
\DeclareMathAlphabet\mathchorus     {T1}{qzc} {m} {n}
\DeclareMathAlphabet\mathrsfso      {U}{rsfso}{m}{n}
\DeclareMathAlphabet\mathfrcal      {T1}{frcursive}{m}{it}
\DeclareFontFamily{T1}{frcursive}{}
\DeclareFontShape{T1}{frcursive}{m}{n}{<->s*[1.44]callig15}{}
\DeclareMathAlphabet\mathfrcal      {T1}{frcursive}{m}{it}

\usepackage{amsmath}
\usepackage{amssymb}
\usepackage{graphicx}
\usepackage[dvipsnames]{xcolor}
\usepackage{soul}
\numberwithin{equation}{section}
\usepackage{array}
\usepackage{mathtools}
\usepackage{dsfont}
\usepackage{mathrsfs}

\usepackage{BOONDOX-uprscr}

\usepackage{tikz}\usetikzlibrary{matrix,fit}
\usepackage{varwidth}
\usepackage{enumerate}
\usepackage{appendix}

\usepackage[margin=1in]{geometry}
\usepackage{nicematrix}

\usepackage[
    backend=bibtex,
    style=alphabetic,
maxbibnames=99,
giveninits=true,
minalphanames=1,
maxalphanames=3,url=false
  ]{biblatex}

\bibliography{SUSYGrassmannian}
\usepackage{mathabx}
\usepackage{empheq}

\usepackage{setspace}

\usepackage{slashed}
\usepackage{upgreek}
\usepackage{appendix}

\usepackage{tabstackengine}

\usepackage{wrapfig}
\usepackage[abs]{overpic}

\usepackage{float}

\fixTABwidth{T}

\newdimen\mytextwidth
\newcommand\rem[2][cyan!40!green]{\noindent\nobreak\hfil\penalty1000\hfilneg
\mytextwidth=\linewidth\advance\mytextwidth by 2mm
\begin{tikzpicture}[baseline=-\the\dimexpr\fontdimen22\textfont2\relax]\node[outer sep=0pt,draw=black,fill=#1,fill opacity=1,text opacity=1,rectangle,rounded corners]{\begin{varwidth}{\mytextwidth}\textcolor{white}{#2}\end{varwidth}};
\end{tikzpicture}\allowbreak
}

\newcommand\whiterem[2][white!]{\noindent\nobreak\hfil\penalty1000\hfilneg
\mytextwidth=\linewidth\advance\mytextwidth by 2mm
\begin{tikzpicture}[baseline=-\the\dimexpr\fontdimen22\textfont2\relax]\node[outer sep=0pt,draw=black,fill=#1,fill opacity=1,text opacity=1,rectangle,rounded corners,line width=1.5pt]{\begin{varwidth}{\mytextwidth}\textcolor{black}{#2}\end{varwidth}};
\end{tikzpicture}\allowbreak
}

\newcommand{\bea}{\begin{equation}}
\newcommand{\eea}{\end{equation}}
\newcommand{\bear}{\begin{eqnarray}}
\newcommand{\eear}{\end{eqnarray}}
\newcommand{\bearr}{\begin{eqnarray*}}
\newcommand{\eearr}{\end{eqnarray*}}

\renewbibmacro{in:}{}

\usepackage{mdframed}

\newbibmacro{string+doi}[1]{
  \iffieldundef{doi}{#1}{\href{http://dx.doi.org/\thefield{doi}}{#1}}}
\DeclareFieldFormat{title}{\usebibmacro{string+doi}{\mkbibemph{#1}}}
\DeclareFieldFormat[article]{title}{\usebibmacro{string+doi}{\mkbibquote{#1}}}

\newmdenv[
  topline=false,
  bottomline=false,
  rightline=false,
  linewidth=2pt,
  skipabove=\topsep,
  skipbelow=\topsep
]{siderules}

\newmdenv[
  topline=false,
  bottomline=false,
  linewidth=2pt,
  skipabove=\topsep,
  skipbelow=\topsep
]{siderulesright}

\makeatletter
\renewcommand{\@seccntformat}[1]{\csname the#1\endcsname.\quad}
\makeatother

\usepackage{setspace}
\usepackage{xpatch}

\makeatletter
\renewcommand{\@chap@pppage}{
  \clear@ppage
  \thispagestyle{plain}
  \if@twocolumn\onecolumn\@tempswatrue\else\@tempswafalse\fi
  \null\vfil
  \markboth{}{}
  {\centering
   \interlinepenalty \@M
   \normalfont
   \MakeUppercase \appendixpagename\par}
  \if@dotoc@pp
    \addappheadtotoc
  \fi
  \vfil\newpage
  \if@twoside
    \if@openright
      \null
      \thispagestyle{empty}
      \newpage
    \fi
  \fi
  \if@tempswa
    \twocolumn
  \fi
}
\makeatother

\definecolor{navycol}{RGB}{100,150,160}
   \definecolor{pinkcol}{RGB}{242,55,55}
   \definecolor{greencol}{RGB}{50,205,50}

   \definecolor{bluecol}{RGB}{30,144,255}

\usepackage{titlesec}

\titleformat*{\section}{\large\bfseries}
\titleformat*{\subsection}{\normalsize\bfseries}
\titleformat*{\subsubsection}{\normalsize\bfseries}
\titleformat*{\paragraph}{\large\bfseries}
\titleformat*{\subparagraph}{\large\bfseries}
\titlespacing{\author}{-5pt}{-5pt}{-5pt}[-5pt]

\makeatletter
\renewcommand\subsubsection{\@startsection{subsubsection}{3}{\z@}
                                     {-3.25ex\@plus -1ex \@minus -.2ex}
                                     {-1.5ex \@plus -.2ex}
                                     {\normalfont\normalsize\bfseries}}
\renewcommand\subsection{\@startsection{subsection}{3}{\z@}
                                     {-3.25ex\@plus -1ex \@minus -.2ex}
                                     {-1.5ex \@plus -.2ex}
                                     {\normalfont\normalsize\bfseries}}                                     
\makeatother

\DeclareFontFamily{U}{solomos}{}
\DeclareErrorFont{U}{solomos}{m}{n}{10}
\DeclareFontShape{U}{solomos}{m}{n}{
  <-> s*[1.1]  gsolomos8r
}{}

\newcommand{\vkappa}{\text{\usefont{U}{solomos}{m}{n}\symbol{'153}}}

   \usepackage{stmaryrd}

\usepackage{tikz}
\usetikzlibrary{arrows.meta}

\usepackage{indentfirst}

\usepackage{tocloft}
\let \savenumberline \numberline
\def \numberline#1{\savenumberline{#1.}}

\usepackage{etoolbox}
\patchcmd{\tableofcontents}{\@starttoc}{\vspace{-0.3cm}\@starttoc}{}{}

\usepackage{accents}
\newcommand\thickbar[1]{\accentset{\rule{.7em}{.8pt}}{#1}}

\newcommand\smallthickbar[1]{\accentset{\rule{.5em}{.8pt}}{#1}}

\usepackage{hyperref}
\hypersetup{
colorlinks=true,
linkcolor=violet,
citecolor=violet,
filecolor=purple,
urlcolor=cyan,
breaklinks=true
}

\begin{document}

\title{\vspace{-1.0cm} \textbf{On First-Order GLSM for Sigma Models}
\vspace{0.5cm}
}

\author{Viacheslav Krivorol$^{\,a,\,b,\,}$\footnote{Emails:
 vkrivorol@itmp.msu.ru, v.a.krivorol@gmail.com}
\\  \vspace{0cm}  \\
{\small $a)$ \emph{Institute for Theoretical and Mathematical Physics,}} \\{\small \emph{Lomonosov Moscow State University, 119991 Moscow, Russia}}\\
{\small $b)$ \emph{Steklov
Mathematical Institute of Russian Academy of Sciences,}} \\{\small \emph{Gubkina str. 8, 119991 Moscow, Russia}}}

\date{}

{\let\newpage\relax\maketitle}

\maketitle

\vspace{0cm}
\textbf{Abstract.} 
We review some recent developments in 1-st order GLSM construction, or so-called Gross-Neveu formalism for sigma models. We recall the general idea behind this framework and describe a 1-st order GLSM data from which the general generalized Gross-Neveu model can be constructed, using $\mathbb{CP}^{n-1}$ sigma model as simplest example. Then, we formulate a general statement that answers the question of which generalized Gross-Neveu models are equivalent to sigma models on compact homogeneous Hermitian spaces of classical groups endowed with the normal metrics.

\newpage

\newpage

\section*{Introduction}
Sigma models are an important class of models in modern theoretical and mathematical physics. They find applications in a wide variety of contexts, from pure mathematics, string theory, etc., to condensed matter physics, spin chain theory, and the like. However, these theories are often highly non-linear, making them difficult to work with. Technically, this is manifested in the fact that the Lagrangian of these models, generally, contains an infinite number of interactions, which makes standard QFT methods (like the perturbation theory w.r.t.~coupling constants) inapplicable. It is possible, of course, to develop some alternatives to perturbation theory (for recent developments see \cite{Alfimov1,Alfimov2}), for example the background field method. The strategy here is to decompose the quantum fields into classical fields (i.e.~solving the equations of motion) and quantum corrections, which we assume to be small. Limiting ourselves to a certain degree of quantum correction in the Lagrangian, we arrive at a field theory with a finite number of interactions. However, this method has a number of disadvantages. In particular, calculations can be extremely difficult and challenging to automate. The choice of background field can be ambiguous, among other things. These difficulties motivate us to find a way to reduce the complexity of the original problem formulation by considering additional information, such as the symmetry of the system.  In this note, we discuss one method for a specific class of models with a rich symmetry structure. It is called the ``Gross-Neveu formalism'', or the ``first-order GLSM formulation''.

\label{sec:preparation}
\section{What do we mean by sigma models?}
Before discussing a formalism, we briefly focus on the general convenient definition of the models under consideration. Let us be given the following geometric data: a Riemann surface\footnote{For the purposes of our paper we always assume $\Sigma$ to be $\mathbb{C}$. But the case of the general Riemann surfaces is also extremely interesting, see \cite{BykovRiemann}.} $\Sigma$ (the worldsheet) with complex coordinate $z$ and a complex manifold $\mathcal{M}$ (the target space) with complex coordinates $\big\{u^i\big\}_{i=1}^{\mathrm{dim}_\mathbb{C}\mathcal{M}}$  
and a Hermitian metric $g = g_{ij}\mathrm{d}u^i\mathrm{d}\smallthickbar{u}^j$. 
Then the sigma model is the field theory on smooth maps $\varphi:\Sigma\xrightarrow[~]{~}\mathcal{M}$ with the action functional
\begin{equation}
\mathcal{S}[\varphi]:=\mathcal{S}[u] = \int_\Sigma\mathrm{d}^2 z\,\Big(g_{ij}(u,\smallthickbar{u})\,\smallthickbar{\partial}u^i\partial\smallthickbar{u}^j\Big)
\end{equation}
where all the coordinate functions are pulled back onto the worldsheet by $\varphi$. Note that the our definition differs from the generally accepted one, see the discussion in \cite{BykovSMAGNM}.
A prototypical example is the $\mathbb{CP}^{n-1}$ sigma model endowed with the Fubini-Study metric,
\begin{equation}
\mathcal{S}[u] = \int_\Sigma \mathrm{d}^2z \Bigg(\frac{\big(1+|u|^2\big)|\smallthickbar{\partial}u|^2 - |\smallthickbar{u}\cdot\smallthickbar{\partial}u|^2}{2\pi\vkappa\big(1+|u|^2\big)^2}\Bigg)\,.
\end{equation}
Here $u$ is the ``vector'' made up of inhomogeneous coordinates $u^1,\ldots,u^{n-1}$, the dot $\cdot$ is the scalar product (contraction of indices), $|u|^2 := \smallthickbar{u}\cdot{u}$ and $\vkappa$ is a coupling constant. 
\section{General idea for the first-order GLSM}
As already mentioned in introduction, the main problem with sigma models is that generally the target space is extremely nonlinear. The idea is to move from a theory with a complicated target space to a gauge theory with a \textit{flat ``phase space''}\footnote{Regarding the pure first-order formulation of sigma models see \cite{LosevMarshakov,GamayunLosevShifman}.}
\begin{equation}
\label{CpAction}
\underbrace{\mathcal{M}}_{\text{standard formulation}}\rightsquigarrow\underbrace{\mathrm{T}^\ast\mathcal{M}}_{\text{1-st order formulation}}\simeq\underbrace{\mathrm{T}^\ast\mathbb{C}^N\sslash \mathcal{G}_\mathbb{C}}_{\text{1-st order GLSM (``Gross-Neveu'')}}
\end{equation}
for some $N$ and a \textit{complex gauge group} $\mathcal{G}_\mathbb{C}$ (by $\sslash$ we mean the Hamiltonian reduction\footnote{More precisely, one actually need to use the so-called GIT quotient if the group action is not free. It differs from Hamiltonian reduction in that some ``bad points'' are thrown out of $\mathrm{T}^\ast\mathbb{C}^N$ before factorization. This makes it possible to ensure that the result is again a smooth symplectic manifold. By $\sslash$ here and hereafter we mean exactly the GIT quotient.}  by $\mathcal{G}_\mathbb{C}$ of $\mathrm{T}^\ast\mathbb{C}^N$ with standard symplectic form). Up to some technical assumptions the manifolds equipped with this representation are called \textit{the quiver varieties} \cite{Nakajima}.
This scheme is something similar to the Olshanetsky-Perelomov trick from the theory of integrable systems \cite{OP}. If such a formulation of the cotangent bundle to the target space exists, our method consists in postulating some action of a special form and then substantiating its equivalence to a sigma model. We do the construction in two steps.

Let us first discuss the kinetic term of such theories. Suppose $u$ and $v$ are vector and covector made up of canonical coordinates and momenta on $\mathrm{T}^\ast\mathbb{C}^N$ respectively and $\mu=\mu(v,u)\in\mathfrak{g}_\mathbb{C}^\ast$ is the moment map of the symplectic reduction w.r.t.~$\mathcal{G}_\mathbb{C}$. Then we can define the following \textit{gauged} $\beta\gamma$-\textit{system} \cite{Nekrasov:2005wg},
\begin{equation}
\label{FreeAction}
\mathcal{S}_0[v,u,A] = \int_\Sigma\mathrm{d}^2 z\Big(v\cdot\smallthickbar{\partial}u+\langle\thickbar{\mathcal{A}},\mu\rangle-\smallthickbar{v}\cdot\partial\smallthickbar{u}-\langle\mathcal{A},\smallthickbar{\mu}\rangle\Big)\,,
\end{equation}
where $A = \mathcal{A}\mathrm{d}z+\thickbar{\mathcal{A}}\mathrm{d}\smallthickbar{z}$ is the $\mathfrak{g}_\mathbb{C}$-valued gauge field on $\Sigma$ (the Lagrange multiplier for the constraint $\mu = 0$) and $\langle\bullet,\bullet\rangle$ is the pairing between $\mathfrak{g}_\mathbb{C}$ and $\mathfrak{g}_\mathbb{C}^\ast$. Hereafter, however, we sometimes think of the $\mu$ as a comoment, i.e.~as an element of $\mathfrak{g}_\mathbb{C}$ which is mapped from $\mathfrak{g}^\ast_\mathbb{C}$ by the Killing metric (which we denote by $\mathrm{Tr}$). 

Next, we introduce interaction into this theory.
To do this, we need to find an other group\footnote{The connection between groups $\mathcal{G}_\mathbb{C}$ and $\widetilde{\mathcal{G}_\mathbb{C}}$ seem to be closely related to the symplectic dual pairs \cite{Weinstein2021LocalSO} and the Howe duality \cite{Howe1989RemarksOC}, see the discussion in the section 4 of \cite{Basile:2023vyg}. It is interesting to clarify this point in the future work.} 
$\widetilde{\mathcal{G}_\mathbb{C}}$ of global symmetries of the holomorphic sector of the free action (\ref{FreeAction}). The action of this group has some comoment map (Noether current) which we denote by $J\in\widetilde{\mathfrak{g}_\mathbb{C}}$. Then, we can couple the holomorphic and anti-holomorphic sectors of the theory (\ref{FreeAction}) by a current-current term with a coupling constant $\vkappa$, i.e.
\begin{equation}
\label{gGN}
\mathcal{S}_{\mathrm{GN}} = \mathcal{S}_0 +2\pi\vkappa \int_\Sigma\mathrm{d}^2 z \,\mathrm{Tr}\big(J\smallthickbar{J}\,\big)\,.
\end{equation}
We call this action \textit{the 1-st order GLSM action}, or \textit{generalized (chiral) Gross-Neveu action}. Actions of this type can be considered as a generalization of chiral Gross-Neveu models \cite{BykovSMAGNM}. The motivation for introducing such an action is that a certain class of sigma models can be identically (at least at classical level) rewritten in this form. Moreover, the number of interaction vertices in such theories is finite, and the gauge field is topological \cite{BykovRiemann} (for a flat worldsheet, this means that there is a gauge $A = 0$).

\section{$\mathbb{CP}^{n-1}$ sigma models}
In this section we discuss the 1-st order GLSM for the simplest particular case $\mathbb{CP}^{n-1}$ \cite{BykovSMAGNM}.
It is well known fact that $\mathrm{T}^\ast\mathbb{CP}^{n-1}\simeq \mathrm{T}^\ast\mathbb{C}^n\sslash\mathbb{C}^\ast$, where $\mathbb{C}^\ast$ is the multiplicative group of nonzero complex numbers. Thus, we have the following data for the 1-st order GLSM construction:
\begin{itemize}
    \item The gauge group $\mathcal{G}_{\mathbb{C}} = \mathbb{C}^\ast$ with the action $u\rightarrow\lambda u$, $v\rightarrow\lambda^{-1}v$, where $\lambda\in\mathbb{C}^\ast$, and the comoment map $\mu = v\cdot u$,
    \item The group of free global symmetry $\widetilde{\mathcal{G}_\mathbb{C}} = \mathbf{GL}(n,\mathbb{C})$ with the action $u\rightarrow h u$, $v\rightarrow vh^{-1}$ , where $h\in \mathbf{GL}(n,\mathbb{C})$, and the current $J = u\otimes v$ which is the $n\times n$ matrix field.
\end{itemize}
By using this data one can construct the generalized Gross-Neveu action (\ref{gGN}). It can be rewritten explicitly as
\begin{equation}
\label{CpGN}
\mathcal{S}_{\mathrm{GN}} = \int_\Sigma\mathrm{d}^2 z \Big(v\cdot\thickbar{D}u-\smallthickbar{v}\cdot D\smallthickbar{u}+2\pi\vkappa\,|u|^2|v|^2\Big)\,,\quad \thickbar{D}u = \smallthickbar{\partial}u+\thickbar{\mathcal{A}}\,u\,.
\end{equation}
The moral is that the action (\ref{CpGN}) classically equivalent to the $\mathbb{CP}^{n-1}$ action
(\ref{CpAction}).
The proof of this statement is the simple exclusion of fields $v$ and $\mathcal{A}$ (and their complex conjugate) from the equations of motion and the fixing the ``inhomogeneous gauge'' $u^n = 1$. 

Note that we can choose in (\ref{CpGN}) an alternative gauge $\mathcal{A} = \thickbar{\mathcal{A}} = 0$. In this gauge the theory is significantly simple and can be seen as 2D analog of $\varphi^4$ action. That's the great thing about the 1-st order GLSM construction. For example, we can recover the one-loop beta function $\beta_\vkappa = -n\vkappa^2+\mathcal{O}(\vkappa^3)$ calculating only 2 simple ``fish type'' diagrams (without any background field method), see the section 6 in \cite{BykovKrivorol}.

\section{Sigma models for compact Hermitian homogeneous spaces}
One can ask, for which sigma models can 1-st order GLSM construction be in principle performed? Unfortunately, this question is quite complex and the answer to it is not known in full generality. However, we can try to answer it for some ``nice'' class of manifolds and metrics. Since it is obvious that symmetry plays a big role in this story, it is reasonable to try to take homogeneous spaces as manifolds. By construction, it is necessary that a Hermitian metric can be introduced on such a manifold, and if possible, this metric should be natural in a certain sense. It is also convenient for simplicity that the global symmetry groups that occur should be compact, semi-simple, and classical. Taking the above into account, we come to the problem of constructing a Gross-Neveu formalism for Hermitian homogeneous spaces of compact type constructed on classical groups. Remarkably, it turns out that the list of such manifolds is quite limited\footnote{Here by orthogonal and symplectic groups we mean the compact forms of the corresponding complex groups.}:
\begin{equation}
\label{Grassmannians}
\frac{\mathbf{U}(n)}{\mathbf{U}(m)\times \mathbf{U}(m-n)}\,,\quad
\frac{\mathbf{O}(n)}{\mathbf{U}(m)\times \mathbf{O}(n-2m)}\,,\quad
\frac{\mathbf{Sp}(2n)}{\mathbf{U}(m)\times \mathbf{U}(2n-2m)}\,,
\end{equation}
and they are usually referred as unitary, orthogonal and symplectic Grassmannians respectively. It is a little less obvious which invariant metrics should be taken on these manifolds. However, it turns out that the so-called \textit{normal metrics} are a good choice. Let us briefly recall the definition. Suppose we have a semi-simple Lie group $\mathcal{G}$ with a subgroup $\mathcal{H}\hookrightarrow \mathcal{G}$ and the corresponding Lie algebras are $\mathfrak{g}$ and $\mathfrak{h}$. Then we can decompose $\mathfrak{g} = \mathfrak{h}\oplus\mathfrak{m}$, where $\mathfrak{m}$ is the orthogonal complement to $\mathfrak{h}$ w.r.t.~Killing metric. We can analogously decompose the corresponding Maurer-Cartan form $\theta = g^{-1}\mathrm{d}g$ as $\theta=\theta_\mathfrak{h}+\theta_\mathfrak{m}$. Then one can introduce the normal metric on the coset $\mathcal{G}/\mathcal{H}$ as $\mathrm{d}s^2_{\text{norm}} = -\mathrm{Tr}\,\theta_\mathfrak{m}^2$. It turns out that this type of metric is the most natural in our context.

Now we are ready to describe the 1-st order GLSM data for manifolds from the specified class. All of this target space cotangent bundles have the symplectic quotient form $\mathrm{T}^\ast\mathrm{Mat}_{m,n}(\mathbb{C})\sslash\mathcal{G}_{\mathbb{C}}$ for different $\mathcal{G}_{\mathbb{C}}$'s that we describe in a moment. We use $U$ and $V$ for a canonical conjugate coordinates on $\mathrm{T}^\ast\mathrm{Mat}_{m,n}(\mathbb{C})$ which are $m\times n$ and $n\times m$ matrices respectively. 
All the groups $\widetilde{\mathcal{G}_{\mathbb{C}}}$ are subgroups in $\mathbf{GL}(n,\mathbb{C})$ and they act as $U\rightarrow h U$, $V\rightarrow Vh^{-1}$. Next, there are three situations corresponding to different Grassmannians in (\ref{Grassmannians}):
\begin{itemize}
    \item Unitary: the gauge group $\mathcal{G}_{\mathbb{C}} = \mathbf{GL}(m,\mathbb{C})$ acts as $U\rightarrow Ug$, $V\rightarrow g^{-1}V$ with the comoment map $\mu = VU$. The global group $\widetilde{\mathcal{G}_{\mathbb{C}}}$ is\footnote{It turns out that it can be reduced to $\mathbf{PSL}(n,\mathbb{C})$.} $\mathbf{GL}(n,\mathbb{C})$ and the corresponding current is $J=UV$.
    \item Orthogonal: the gauge group $\mathcal{G}_{\mathbb{C}} = \mathbf{GL}(m,\mathbb{C})\rtimes\mathrm{Mat}_{m}^{\text{symm}}$, where $\mathrm{Mat}_{m}^{\text{symm}}$ is the abelian group of symmetric $m\times m$ matrices, they act as $U\rightarrow Ug$ and $V\rightarrow g^{-1}V+qU^t$, $q\in \mathrm{Mat}_{m}^{\text{symm}}$. It is convenient to think that we have two comoment maps corresponding to the first and second groups of the semidirect product: $\mu_1 = VU$ and $\mu_2 = U^tU$. The global group $\widetilde{\mathcal{G}_{\mathbb{C}}}$ is $\mathbf{SO}(n,\mathbb{C})$ and the corresponding current is $J=UV-(UV)^t$.
    \item Symplectic: It repeats the previous case almost verbatim with minor modifications. 
    Let's just say that $\mathcal{G}_{\mathbb{C}} = \mathbf{GL}(m,\mathbb{C})\rtimes\mathrm{Mat}_{m}^{\text{skew-symm}}$ and $\widetilde{\mathcal{G}_{\mathbb{C}}}$ is $\mathbf{Sp}(2n,\mathbb{C})$.
\end{itemize}
In fact, everything is ready for the formulation of the main statement \cite{BykovKrivorol}: the generalized Gross-Neveu models, constructed from the above described 1-st order GLSM data, are equivalent to the sigma models which target spaces are corresponding unitary, orthogonal and symplectic Grassmannians (\ref{Grassmannians}), equipped with the normal metric. 

A generalization of this statement to the supersymmetric case is given in \cite{BykovFermions} for $\mathbb{CP}^{n-1}$ and in \cite{BykovKrivorol2} for Grassmannians (\ref{Grassmannians}). Deformations of the normal (round) metric within described formalism for $\mathbb{CP}^1$ were studied in \cite{BykovPribytok}, see also \cite{Pribytok:2024lej}. Some representation theory applications of this formalism for mechanics on $\mathbb{CP}^{n-1}$ can be find in \cite{BykovSmilga}. For discussion of relation between generalized Gross-Neveu models and nilpotent orbits of complex Lie groups see \cite{Bykov:2019vkf,Bykov2}.

\label{sec:acknowledgement}
\section*{Acknowledgement}
I would like to thank Andrew Kuzovchikov, Mikhail Markov, and especially Dmitri Bykov for useful discussions, comments on this work, and inspiration.
This work is supported by the Russian Science Foundation grant № 22-72-10122.

\setstretch{0.8}
\setlength\bibitemsep{5pt}
\printbibliography

@article{BykovKrivorol,
    author = "Bykov, Dmitri and Krivorol, Viacheslav",
    title = "{Grassmannian Sigma Models}",
    eprint = "2306.04555",
    archivePrefix = "arXiv",
    primaryClass = "hep-th",
    doi = "10.4310/ATMP.241028230215",
    journal = "Adv. Theor. Math. Phys.",
    volume = "28",
    pages = "3",
    year = "2024"
}

@article{BykovRiemann,
    author = "Bykov, Dmitri",
    title = "{Integrable sigma models on Riemann surfaces}",
    eprint = "2202.12805",
    archivePrefix = "arXiv",
    primaryClass = "hep-th",
    doi = "10.1103/PhysRevD.107.085015",
    journal = "Phys. Rev. D",
    volume = "107",
    number = "8",
    pages = "085015",
    year = "2023"
}

@article{BykovSMAGNM,
    author = "Bykov, Dmitri",
    title = "{Sigma models as Gross\textendash{}Neveu models}",
    eprint = "2106.15598",
    archivePrefix = "arXiv",
    primaryClass = "hep-th",
    doi = "10.1134/S0040577921080018",
    journal = "Teor. Mat. Fiz.",
    volume = "208",
    number = "2",
    pages = "165--179",
    year = "2021"
}

@article{OP,
    author = "Ol'shanetskii, M. A. and Perelomov, A. M.",
    title = "{The Toda chain as a reduced system}",
    doi = "10.1007/BF01047139",
    journal = "Theoretical and Mathematical Physics",
    volume = "45",
    pages = "843--854",
    year = "1980"
}

@article{Nakajima,
    author = "Nakajima, Hiraku",
    title = "{Instantons on ALE spaces, quiver varieties, and Kac-Moody algebras}",
    doi = "10.1215/S0012-7094-94-07613-8",
    journal = "Duke Math. J.",
    volume = "76",
    number = "2",
    pages = "365--416",
    year = "1994"
}

@article{Bykov:2019vkf,
    author = "Bykov, Dmitri V.",
    title = "{Flag Manifold Sigma Models and Nilpotent Orbits}",
    eprint = "1911.07768",
    archivePrefix = "arXiv",
    primaryClass = "hep-th",
    reportNumber = "LMU-ASC 37/20, MPP-2020-174",
    doi = "10.1134/S0081543820030062",
    journal = "Proc. Steklov Inst. Math.",
    volume = "309",
    number = "1",
    pages = "78--86",
    year = "2020"
}

@article{LosevMarshakov,
    author = "Losev, Andrei S. and Marshakov, Andrei and Zeitlin, Anton M.",
    title = "{On first order formalism in string theory}",
    eprint = "hep-th/0510065",
    archivePrefix = "arXiv",
    reportNumber = "ITEP-TH-59-05, FIAN-TD-16-05",
    doi = "10.1016/j.physletb.2005.12.010",
    journal = "Phys. Lett. B",
    volume = "633",
    pages = "375--381",
    year = "2006"
}

@article{GamayunLosevShifman,
    author = "Gamayun, Oleksandr and Losev, Andrei and Shifman, Mikhail",
    title = "{First-order formalism for \ensuremath{\beta} functions in bosonic sigma models from supersymmetry breaking}",
    eprint = "2312.01885",
    archivePrefix = "arXiv",
    primaryClass = "hep-th",
    reportNumber = "FTPI-MINN-23-23, UMN-TH-4228/23",
    doi = "10.1103/PhysRevD.110.025017",
    journal = "Phys. Rev. D",
    volume = "110",
    number = "2",
    pages = "025017",
    year = "2024"
}

@article{BykovKrivorol2,
    author = "Bykov, Dmitri and Krivorol, Viacheslav",
    title = "{Supersymmetric Grassmannian Sigma Models in Gross-Neveu Formalism}",
    eprint = "2407.20423",
    archivePrefix = "arXiv",
    primaryClass = "hep-th",
    doi = "10.48550/arXiv.2407.20423",
    year = "2024"
}

@article{BykovFermions,
    author = "Bykov, Dmitri",
    title = "{The $\mathsf{CP^{n-1}}$-model with fermions: a new look}",
    eprint = "2009.04608",
    archivePrefix = "arXiv",
    primaryClass = "hep-th",
    reportNumber = "MPP-2020-166, LMU-ASC-35/20",
    doi = "10.4310/ATMP.2022.v26.n2.a2",
    journal = "Adv. Theor. Math. Phys.",
    volume = "26",
    pages = "2",
    year = "2022"
}

@article{BykovPribytok,
    author = "Bykov, Dmitri and Pribytok, Anton",
    title = "{Supersymmetric deformation of the $ \mathbb{CP}^{1} $ model and its conformal limits}",
    eprint = "2312.16396",
    archivePrefix = "arXiv",
    primaryClass = "hep-th",
    doi = "10.4310/ATMP.241119040718",
    journal = "Adv. Theor. Math. Phys.",
    volume = "28",
    pages = "8",
    year = "2024"
}

@inproceedings{Pribytok:2024lej,
    author = "Pribytok, Anton",
    title = "{Superdeformed $\mathbb{CP}$ $\sigma$-model equivalence}",
    booktitle = "{International Workshop Supersymmetries and Quantum Symmetries}",
    eprint = "2412.00670",
    archivePrefix = "arXiv",
    primaryClass = "hep-th",
    doi = "10.48550/arXiv.2412.00670",
    year = "2024"
}

@article{BykovSmilga,
    author = "Bykov, Dmitri and Smilga, Andrei",
    title = "{Monopole harmonics on $\mathbb{CP}^{n-1}$}",
    eprint = "2302.11691",
    archivePrefix = "arXiv",
    primaryClass = "hep-th",
    doi = "10.21468/SciPostPhys.15.5.195",
    journal = "SciPost Phys.",
    volume = "15",
    number = "5",
    pages = "195",
    year = "2023"
}

@article{Nekrasov:2005wg,
    author = "Nekrasov, Nikita A.",
    title = "{Lectures on curved beta-gamma systems, pure spinors, and anomalies}",
    eprint = "hep-th/0511008",
    archivePrefix = "arXiv",
    reportNumber = "IHES-P-05-35, ITEP-TH-58-05",
    doi = "10.48550/arXiv.hep-th/0511008",
    year = "2005"
}

@article{Basile:2023vyg,
    author = "Basile, Thomas and Joung, Euihun and Oh, TaeHwan",
    title = "{Manifestly covariant worldline actions from coadjoint orbits. Part I. Generalities and vectorial descriptions}",
    eprint = "2307.13644",
    archivePrefix = "arXiv",
    primaryClass = "hep-th",
    doi = "10.1007/JHEP01(2024)018",
    journal = "JHEP",
    volume = "01",
    pages = "018",
    year = "2024"
}

@article{Bykov2,
    author = "Bykov, Dmitri",
    title = "{Sigma models as Gross\textendash{}Neveu models. II}",
    eprint = "2310.15394",
    archivePrefix = "arXiv",
    primaryClass = "hep-th",
    doi = "10.1134/S0040577923120048",
    journal = "Theor. Math. Phys.",
    volume = "217",
    number = "3",
    pages = "1842--1854",
    year = "2023"
}

@article{Weinstein2021LocalSO,
  title={The local structure of Poisson manifolds},
  author={Alan J. Weinstein},
    doi = "10.4310/jdg/1214437787",
  journal={Journal of differential geometry},
  year={1983},
  volume={18},
  pages={523–557},
  url={https://projecteuclid.org/journals/journal-of-differential-geometry/volume-18/issue-3/The-local-structure-of-Poisson-manifolds/10.4310/jdg/1214437787.full}
}

@article{Howe1989RemarksOC,
  title={Remarks on classical invariant theory},
  author={Roger E. Howe},
    doi = "10.1090/S0002-9947-1989-0986027-X",
  journal={Transactions of the American Mathematical Society},
  year={1989},
  volume={313},
  pages={539-570},
  url={https://api.semanticscholar.org/CorpusID:43706152}
}

@article{Alfimov1,
    author = "Alfimov, Mikhail and Litvinov, Alexey",
    title = "{On loop corrections to integrable 2D sigma model backgrounds}",
    eprint = "2110.05418",
    archivePrefix = "arXiv",
    primaryClass = "hep-th",
    doi = "10.1007/JHEP01(2022)043",
    journal = "JHEP",
    volume = "01",
    pages = "043",
    year = "2022"
}

@article{Alfimov2,
    author = "Alfimov, Mikhail and Kalinichenko, Ivan and Litvinov, Alexey",
    title = "{On \ensuremath{\beta}-function of N = 2 supersymmetric integrable sigma-models}",
    eprint = "2311.14187",
    archivePrefix = "arXiv",
    primaryClass = "hep-th",
    doi = "10.1007/JHEP05(2024)297",
    journal = "JHEP",
    volume = "05",
    pages = "297",
    year = "2024"
}
\end{document}